\documentclass[twoside,12pt]{article}  %%% 这是 LaTeX 格式文件必须的指令，不可缺少

\usepackage{indentfirst}  %% 节标题后第一行段缩进
\usepackage{bm}    %%%%%  排印数学符号黑斜体时调用此宏包
\usepackage{graphicx}  %%%% 文中插入 eps 图形时调用此宏包，插图方法 1
\usepackage{color}
\begin{document}

%-------------------  First Head  -----------------------------------------
%\thispagestyle{empty} \vspace*{0.8cm}\hbox
%to\textwidth{\vbox{\hfill\huge\sf PHYSICAL REVIEW E\hfill}}
%%\par\noindent\rule[3mm]{\textwidth}{0.2pt}\hspace*{-\textwidth}\noindent
%\rule[2.5mm]{\textwidth}{0.2pt}

%=================== Text begin here =============================================

\begin{center}
%\LARGE\bf \textcolor{blue}{Small disturbances induce ultra-sensitive vibrational resonance: transient behavior}
\LARGE\bf Ultra-sensitive vibrational resonance induced by small disturbances
\end{center}

\begin{center}
Shangyuan Li $^{\rm 1}$, Zhongqiu Wang $^{\rm 2}$, Jianhua Yang $^{\rm 1 *}$, \ \ Miguel A. F. Sanju\'an $^{\rm 3}$, Shengping Huang $^{\rm 1}$, Litai Lou$^{\rm 1}$
\end{center}

\begin{center}
\begin{footnotesize} \sl
${}^{\rm 1}$ Jiangsu Key Laboratory of Mine Mechanical and Electrical Equipment, School of Mechatronic Engineering, China University of Mining and Technology, Xuzhou 221116, Jiangsu, People's Republic of China\\
${}^{\rm 2}$ School of Computer Science and Technology, China University of Mining and Technology, Xuzhou 221116, Jiangsu, People's Republic of China\\
${}^{\rm 3}$ Nonlinear Dynamics, Chaos and Complex Systems Group, Departamento de F\'isica, Universidad Rey Juan Carlos, Tulip\'an s/n, 28933 M\'ostoles, Madrid, Spain\\
${}^{\rm *}$ Corresponding author. E-mail address:jianhuayang@cumt.edu.cn.\\

\end{footnotesize}
\end{center}
%\begin{center}
%\footnotesize (MS received X XX XXXX; revised X XX XXXX)
%\end{center}
\vspace*{2mm}

\begin{center}
\begin{minipage}{15.5cm}
\parindent 20pt\footnotesize
We have found two kinds of ultra-sensitive vibrational resonance in coupled nonlinear systems. It is particularly worth pointing out that this ultra-sensitive vibrational resonance is a transient behavior caused by transient chaos. Considering long-term response, the system will transform from transient chaos to periodic response. The pattern of vibrational resonance will also transform from ultra-sensitive vibrational resonance to conventional vibrational resonance. This article focuses on the transient ultra-sensitive vibrational resonance phenomenon. It is induced by a small disturbance of the high-frequency excitation and the initial simulation conditions, respectively. The damping coefficient and the coupling strength are the key factors to induce the ultra-sensitive vibrational resonance. By increasing these two parameters, the vibrational resonance pattern can be transformed from an ultra-sensitive vibrational resonance to a conventional vibrational resonance. The reason for different vibrational resonance patterns to occur lies in the state of the system response. The response usually presents transient chaotic behavior when the ultra-sensitive vibrational resonance appears and the plot of the response amplitude versus the controlled parameters shows a highly fractalized pattern. When the response is periodic or doubly-periodic, it usually corresponds to the conventional vibrational resonance. The ultra-sensitive vibrational resonance not only occurs at the excitation frequency, but it also occurs at some more nonlinear frequency components. The ultra-sensitive vibrational resonance as a transient behavior and the transformation of vibrational resonance patterns are new phenomena in coupled nonlinear systems.
\end{minipage}
\end{center}

\begin{center}
\begin{minipage}[t]{2.3cm}
{\bf Keywords}:
\end{minipage}
\begin{minipage}[t]{13.1cm}
vibrational resonance, coupled systems, small disturbance, chaotic response
\end{minipage}\par\vglue8pt
\end{center}

\newpage
{\noindent \bf The discovery of relevant nonlinear dynamical phenomena, including chaos, stochastic resonance, coherence resonance, vibrational resonance, and others, has consistently made the investigation of the intricate behavior of nonlinear systems a thriving area of research. The coupling of a system often makes the systems response more complex, and certainly there are numerous nonlinear coupled systems in science and engineering. Furthermore, these systems exhibit interesting and unimaginable dynamical response phenomena under different perturbations. The magic of the nonlinear systems response lies in the frequent occurrence of unpredictable phenomena beyond conventional ones. Consider vibrational resonance as an illustration. When a nonlinear system is excited by two signals with distinct time scales, its response can manifest various phenomena, including conventional vibrational resonance, aperiodic vibrational resonance, subharmonic and superharmonic vibrational resonance, ghost vibrational resonance, entropic vibrational resonance, and even logical vibrational resonance. The effects of the above different types of vibrational resonance gradually change with the variation of the control parameters. Specifically, the relative relationship between the indices for measuring vibrational resonance and the control parameters has a strong regularity, and their relationship can be scaled using a determined curve. Beyond the conventional vibrational resonance mentioned above, herein, we focus on ultra-sensitive vibrational resonance in a coupled system subjected to two harmonic excitations. Interestingly, we find that the ultra-sensitive vibrational resonance is a transient but not a long-term dynamical behavior. The ultra-sensitive vibrational resonance can occur at the excitation frequency and some other nonlinear frequencies, and it is induced by a very small disturbance. The small disturbance may belong to the excitation or the initial conditions of the simulation. The very reason of the appearance of the ultra-sensitive vibrational resonance is the presence of regions with transient chaotic in the system response. Furthermore, an increase of the coupling and damping parameters leads to the transformation of different vibrational resonance patterns. Ultrasensitive vibrational resonance is worth studying as a transient phenomenon caused by very small perturbations.}

\newpage
\section{Introduction}
Vibrational resonance has been a research topic of interest to scholars in the field of nonlinear dynamics in recent years. Since this phenomenon was discovered by Landa and McClintock [1], the theoretical framework of vibrational resonance has been continuously improved. It has gone through stages such as relying on pure numerical simulation analysis, analytical researches [2-5], experimental researches [6-9], as well as research on diverse applications [10-14]. The nonlinear models of vibrational resonance extend from simple ordinary differential equations to delayed equations [15-18], fractional-order equations [19-24], complex networks [25-29], discrete dynamical systems [30], laser systems [6, 31], quantum systems [32, 33], position-dependent mass systems [34], etc. The types of vibrational resonance have expanded from the conventional vibrational resonance to aperiodic vibrational resonance [35, 36], ultra-sensitive vibrational resonance [37], superharmonic and subharmonic vibrational resonance [38, 39], ghost vibrational resonance [40, 41], rescaled vibrational resonance [42, 43], logical vibrational resonance [44-47], entropic vibrational resonance [48, 49], and so on. \\
\indent The analysis of the related characteristics of coupled systems has also received considerable attention in the past years. The coupling enables us to understand much about many natural processes. Sarkar and Ray considered coupled bistable oscillators in which both vibrational resonance and vibrational antiresonance occur under adjusting the high-frequency excitation [50]. Yao and Zhan studied the weak low-frequency signal transmission by using the vibrational resonance method in one-way coupled bistable systems subjected to high-frequency signals [51]. Gandhimathi et al. numerically studied the effect of amplitude modulated force in two coupled overdamped anharmonic oscillators [52, 53]. Asir examined the existence of multiple vibrational resonance and antiresonance in coupled overdamped oscillators and illustrated the influence of coupling strength on the interaction of oscillators [54]. Rajasekar and Murali considered a two coupled Duffing-van der Pol oscillators with a nonlinear coupling and studied the influence of coupling strength on the response [55]. Moukam-Kakmeni et al. analyzed the dynamics of a system consisting of two coupled Duffing oscillators [56]. Qian and Yan studied the two degree-of-freedom nonlinear coupled Duffing equation under external single excitation or dual-excitations [57]. Kenfack presented the results of an investigation of two identical coupled double-well Duffing oscillators subjected to a periodically driven force [58]. Yao et al. investigated the resonance behaviour in a system composed by n-coupled Duffing oscillators where only the first oscillator is driven by a periodic force [59]. More and more investigations have revealed that choosing an appropriate coupling is a very effective way to control the degree of resonance.\\
\indent In all the above studies, the system response is deterministic. However, in many cases, the system response exhibits chaotic characteristics, and even undergoes changes from transient chaos to periodic response. Transient chaos is also an very interesting phenomenon and has been extensively studied in many works. Kantz and Grassberger observed the transient chaos in a deterministic map [60]. They pointed out that the transient chaos is related to a repeller. Paula and Savi analysed the chaotic phenomenon in the nonlinear pendulum of the experiment, indicating that it has a transient chaotic response [61]. The transient chaos is also related to a fractal structure, tending to be eliminated in the periodic stable state. As a matter of fact the true nature is the existence of a non-attracting chaotic set (chaotic saddle) in phase space.\\
\indent For the ultra-sensitive vibrational resonance in a Duffing system [37], within certain parameter ranges, very small changes in the high-frequency excitation can cause severe fluctuations in the system response amplitude. This is a new kind of vibrational resonance. In this paper, we will show the appearance of the ultra-sensitive vibrational resonance in new coupled nonlinear systems, and we will conduct a more in-depth study on ultra-sensitive vibrational resonance and explore some new phenomena. As far as we know, due to the existence of more equilibrium states in coupled nonlinear systems and the influence of multiple parameters, the output of coupled systems is generally more complex. We will study a class of coupled nonlinear systems, and discuss which small disturbances will cause the ultra-sensitive vibrational resonance phenomenon. Of course, we are not limited to the high-frequency disturbance here, but will also explore new disturbance factors. At the same time, we also need to investigate the reasons that lead to the transformation of the system response from conventional vibrational resonance to ultra-sensitive vibrational resonance. Importantly, we will first show that the ultra-sensitive vibrational resonance is directly related to the presence of transient chaos and as a consequence it is only a transient behavior.\\
\indent The outline of the paper is as follows. In Sec.~2, we will present the ultra-sensitive vibrational resonance induced by different small disturbances. One is induced by a high-frequency signal, and another one by disturbing the initial conditions in our numerical simulations. This section will also show the transient characteristics of the ultra-sensitive response. In Sec.~3, we discuss the transformation between the ultra-sensitive vibrational resonance and the conventional vibrational resonance. The influences of coefficients, especially the damping coefficient and the coupling strength of the system, on the response pattern will be mainly considered. In Sec.~4, the reason for the ultra-sensitive vibrational resonance will be revealed by the Poincar\'{e} section and phase trajectory. In the last section, the main conclusions of the paper will be given.

\section{Ultra-sensitive vibrational resonance induced by small disturbances}
In this section, we will present the ultra-sensitive vibrational resonance induced by using a small disturbance to the high-frequency signal and to the initial conditions of the numerical simulation, respectively.
\subsection{Ultra-sensitive vibrational resonance induced by a small high-frequency disturbance}
We consider the coupled nonlinear system described by
\begin{equation}
\left\{ \begin{array}{l}
 {{\ddot x}_1} + 2{\gamma _1}{{\dot x}_1} - \omega _1^2{x_1} + {\beta _1}x_1^3 - 2{g_1}{x_2} = {F_1} + {F_2} \\
 {{\ddot x}_2} + 2{\gamma _2}{{\dot x}_2} - \omega _2^2{x_2} + {\beta _2}x_2^3 - 2{g_2}{x_1} = {F_2} \\
 \end{array} \right.
\end{equation}
where $\gamma_1$ and $\gamma_2$ are damping coefficients, $\omega_1$, $\omega_2$, $\beta_1$, $\beta_2$ are the coefficients of the linear terms and nonlinear terms, respectively, $g_1$ and $g_2$ are coupling coefficients which indicate the coupling strength of the system. The kind of coupling is called as nearest neighbor coupling, which is widely used in the literature [62-66]. For the excitations, $F_1=A\cos\omega t$, $F_2=B\cos\Omega t$, $A$ and $\omega$, $B$ and $\Omega$ are the amplitude and frequency of the low-frequency and high-frequency signals, respectively, and $\omega  \ll \Omega $. In the former works on vibrational resonance, the high-frequency signal is an auxiliary signal to induce the conventional vibrational resonance. In this section, we first would like to view it as a very small high-frequency disturbance. In fact, Eq.~(1) was investigated once by Sarkar and Ray [50], and they found the conventional vibrational resonance and vibrational antiresonance in the system.\\
\indent The coupled nonlinear oscillators can be rewritten as four first-order differential equations, as shown in Eq. (2)\\
\begin{equation}
\left\{
\begin{array}{l}
\frac{dx_1}{{dt}} = y_1 ,\\
\frac{dy_1}{{dt}} = -2{\gamma _1}{y_1} + \omega _1^2{x_1} - {\beta _1}x_1^3 + 2{g_1}{x_2} + {F_1} + {F_2} , \\
\frac{dx_2}{{dt}} = y_2 ,\\
\frac{dy_2}{{dt}} = -2{\gamma _2}{y_3} + \omega _2^2{x_2} - {\beta _1}x_1^3 + 2{g_2}{x_1} + {F_2} .\\
\end{array}
\right.
\end{equation}
We use the fourth-order Runge-Kutta algorithm to numerically find the solutions of the equations. For solving the discrete time series $x_1(t)$, the algorithm is given in Eq.~(3)\\
		\begin{equation}
		\left\{
		\begin{array}{l}
		{r_1} = y_1( n ) ,\\
		{k_1} = -2{\gamma _1}{r_1} + \omega _1^2{x_1}( n ) - {\beta _1}x_1^3( n ) + 2{g_1}{x_2}( n ) + F_1( n ) + F_2( n ) ,\\
		{r_2} = y_1( n ) +{k_1}h/2 ,\\
		{k_2} = -2{\gamma _1}{r_2} + \omega _1^2({x_1}( n )+{r_1}h/2) - {\beta _1}({x_1}( n )+{r_1}h/2)^3 + 2{g_1}({x_2}( n )\\\qquad+{s_1}h/2) + F_1( n ) + F_2( n ) ,\\		
		{r_3} = y_1( n ) +{k_2}h/2 ,\\
		{k_3} = -2{\gamma _1}{r_3} + \omega _1^2({x_1}( n )+{r_2}h/2) - {\beta _1}({x_1}( n )+{r_2}h/2)^3 + 2{g_1}({x_2}( n )\\\qquad+{s_2}h/2) + F_1( n+1 ) + F_2( n+1 ) ,\\
		{r_4} = y_1( n ) +{k_3}h/2 ,\\
		{k_4} = -2{\gamma _1}{r_4} + \omega _1^2({x_1}( n )+{r_4}h/2) - {\beta _1}({x_1}( n )+{r_3}h/2)^3 + 2{g_1}({x_2}( n )\\\qquad+{s_3}h/2) + F_1( n+1 ) + F_2( n+1 ) ,\\
		y_1( n+1 ) = y_1( n ) + ( {k_1} +2{k_2} + 2{k_3} + {k_4} )h/6 ,\\
		x_1( n+1 ) = x_1( n ) + ( {r_1} +2{r_2} + 2{r_3} + {r_4} )h/6 .\\				
		\end{array}
		\right.
		\end{equation}
\indent For solving the discrete time series $x_2(t)$, the algorithm is shown in Eq.~(4)\\
		\begin{equation}
		\left\{
		\begin{array}{l}
		{s_1} = y_2( n ) ,\\
		{d_1} = -2{\gamma _2}{s_1} + \omega _2^2{x_2}( n ) - {\beta _2}x_2^3( n ) + 2{g_2}{x_1}( n ) + F_1( n ) + F_2( n ) ,\\
		{s_2} = y_2( n ) +{d_1}h/2 ,\\
		{d_2} = -2{\gamma _2}{s_2} + \omega _2^2({x_2}( n )+{s_1}h/2) - {\beta _2}({x_2}( n )+{s_1}h/2)^3 + 2{g_2}({x_1}( n )\\\qquad+{r_1}h/2) + F_1( n ) + F_2( n ) ,\\		
		{s_3} = y_2( n ) +{d_2}h/2 ,\\
		{d_3} = -2{\gamma _2}{s_3} + \omega _2^2({x_2}( n )+{s_2}h/2) - {\beta _2}({x_2}( n )+{s_2}h/2)^3 + 2{g_2}({x_1}( n )\\\qquad+{r_2}h/2) + F_1( n+1 ) + F_2( n+1 ) ,\\
		{s_4} = y_2( n ) +{d_3}h/2 ,\\
		{d_4} = -2{\gamma _2}{s_4} + \omega _2^2({x_2}( n )+{s_3}h/2) - {\beta _2}({x_2}( n )+{s_3}h/2)^3 + 2{g_2}({x_1}( n )\\\qquad+{r_3}h/2) + F_1( n+1 ) + F_2( n+1 ) ,\\
		y_2( n+1 ) = y_2( n ) + ( {d_1} +2{d_2} + 2{d_3} + {d_4} )h/6 ,\\
		x_2( n+1 ) = x_2( n ) + ( {s_1} +2{s_2} + 2{s_3} + {s_4} )h/6 .\\				
		\end{array}
		\right.
		\end{equation}
In Eqs.~(3) and (4), $F_1(n)$ and $F_2(n)$ are sequences of the low-frequency signal and the high-frequency signal, respectively, where $h$ is the time step.\\
\indent When studying the ultra-sensitive vibrational resonance, analytical methods of the system response will fail, so we use numerical simulations for our research. For the numerical simulation, we use the response amplitude to measure the response of the nonlinear system at a harmonic component. Herein, to quantify the nonlinear response to the low-frequency signal, we calculate the sine and cosine Fourier coefficients of the response $x_1(t)$ or $x_2(t)$ at a series of frequency components, i.e.,
\begin{equation}
\left\{ \begin{array}{l}
 {C_{si}}(m\omega ) = \frac{2}{{nT}}\int_0^{nT} {{x_i}(t)\sin (m\omega t)dt} , \\
 {C_{ci}}(m\omega ) = \frac{2}{{nT}}\int_0^{nT} {{x_i}(t)\cos (m\omega t)dt} , \\
 \end{array} \right.\quad i = 1,\;2.
\end{equation}
Then, the response amplitude at the low-frequency $m\omega$ is given by
\begin{equation}
Q_i = \frac{{\sqrt {C_{si}^2 + C_{ci}^2} }}{A}, i=1,2.
\end{equation}
\indent We concentrate on both vibrational resonance occurring at the excitation frequency $\omega$ and nonlinear vibrational resonance at other nonlinear frequencies in the following analysis, which is a very important variant of the conventional vibrational resonance. As a result, $m$ in Eq.~(5) can be an integer or a fractional  number. In the following analysis, we consider $m=1/3$, $1/2$, $1$, $2$, $3$, respectively. Certainly, $m$ can also take other different values. Compared to the conventional vibrational resonance and nonlinear vibrational resonance, the ultra-sensitive vibrational resonance occurs at other nonlinear frequencies. It is the reason for extending the computation of the response amplitude $Q_i$ in Eqs.~(5) and (6).\\
\indent Firstly, as shown in Fig.~1, it displays the presence of the ultra-sensitive vibrational resonance in the coupled systems described by Eq.~(1). The response amplitude $Q_1$ fluctuates over a large range even though a very small interval of the amplitude $B$ is chosen, as presented in Fig.~1(a). This is the typical pattern of the ultra-sensitive vibrational resonance. We further narrow down the scope of $B$ and the step of $B$ in Fig.~1(b), the ultra-sensitive vibrational resonance still occurs. Figure 1 demonstrates that for a high-frequency disturbance with a very small amplitude may bring a strong resonance and the response of the system is very sensitive to the high-frequency disturbance. The response curve presents very sharp and narrow peaks, which may make the two adjacent peaks arbitrarily close. In Fig.~1(c), we compare two response time series corresponding to two different cases, i.e., the presence and absence of the small high-frequency. Apparently, the response of $x_1$ changes from irregular response to periodic response. Interestingly, a very small high-frequency disturbance makes the system response to quickly transform from irregular to periodic. As far as we know, this fact has not been reported in previous investigations.\\
\begin{figure}[h!]
\centering
\includegraphics[width=6in]{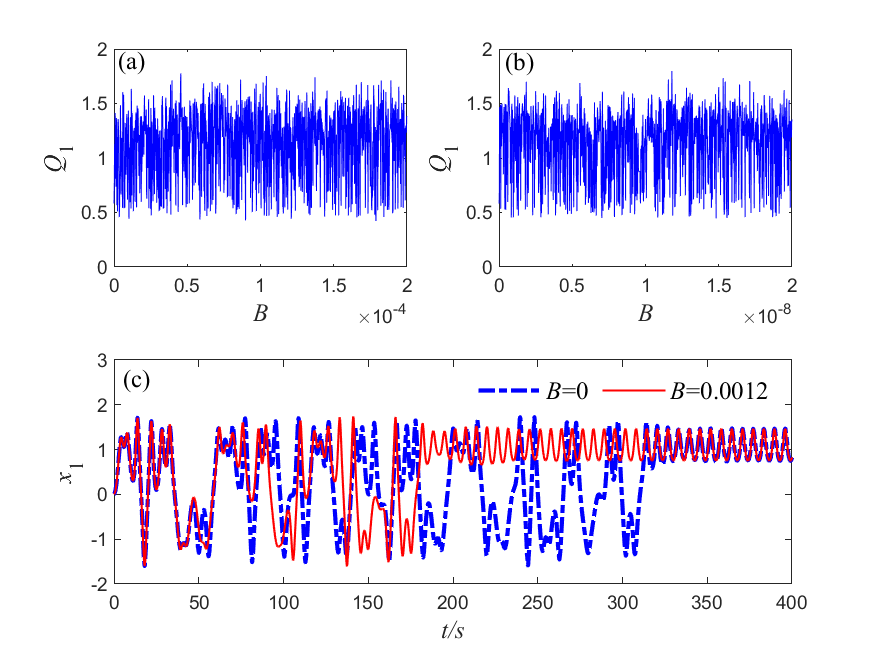}
\parbox{17cm}{\small {\bf FIG.~1.} A very small disturbance of the amplitude of the high-frequency signal induces ultra-sensitive vibrational resonance of $x_1$ at the frequency $\omega$. (a) The response amplitude $Q_1$ versus the amplitude $B$ in a very small range, the step of $B$ is $\Delta B=2 \times 10^{-7}$. (b) The response amplitude $Q_1$ versus the disturbance amplitude $B$ in a much smaller range, the step of $B$ is $ \Delta B=2 \times 10^{-11}$. (c) The response of the system in absence and presence of a small high-frequency disturbance. The parameters are $A = 0.25$, $\omega = 1$, $\Omega = 10$, ${\gamma _1} = {\gamma _2} = 0.05$, $\omega _1^2 = \omega _2^2 = 0.8$, ${g_1} = {g_2} = 0.05$, ${\beta _1} = {\beta _2} = 0.7$.}\label{fig1}
\end{figure}
\indent Actually, $x_1$ and $x_2$ are mutually coupled. To further compare the difference between $x_1$ and $x_2$, we give Fig.~2. In the system model, $x_1$ is disturbed by both low-frequency and high-frequency signals, while $x_2$ acted only by the high-frequency signal. The amplitudes and frequency components of $x_1$ and $x_2$ may have a huge difference, as shown in Fig.~2. After a long time, the transient response turns to the steady response. Specifically, the irregular motion stabilizes in periodic motion. In the first $200$ seconds, from both time series and the spectrum plots, it illustrates that the system response exhibits transient chaotic motion, while finally falls into periodic motion. It is a transformation from transient chaos to periodic motion. This phenomenon has also been widely reported on numerical or experimental results [67, 68]. In this paper, we mainly focus on the ultra-sensitive vibrational resonance that occurs in the transient chaotic region. Hence, in the following analysis, we always use the time series in the first $200$ seconds for calculation.\\
\begin{figure}[h!]
\centering
\includegraphics[width=6in]{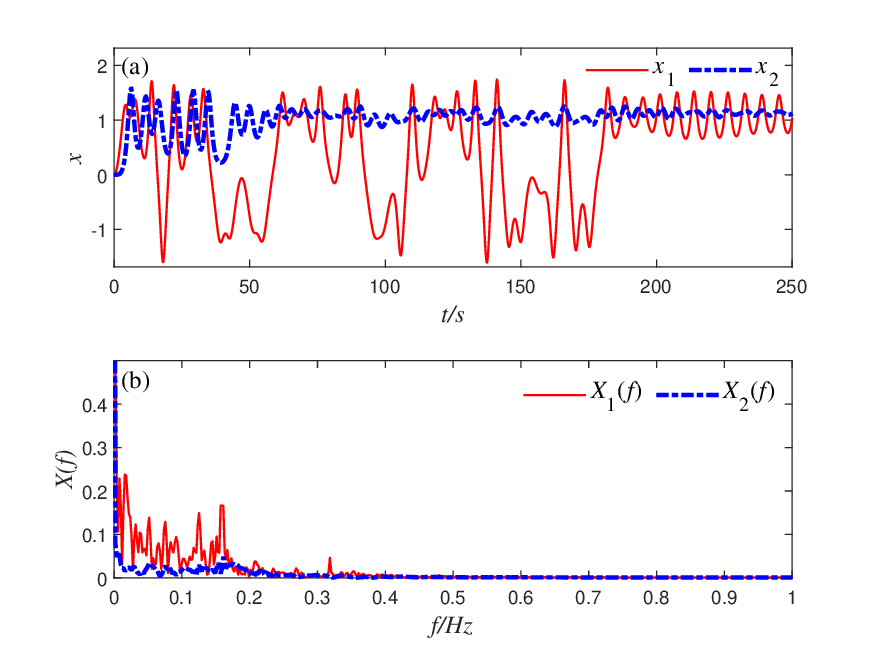}
\parbox{17cm}{\small {\bf FIG.~2.} The response of the system under the high-frequency disturbance. (a) The time domain waveform of the response, (b) The spectrum of the response. The parameters are $A = 0.25$, $\omega = 1$, $B=0.0012$, $\Omega = 10$, ${\gamma _1} = {\gamma _2} = 0.05$, $\omega _1^2 = \omega _2^2 = 0.8$, ${g_1} = {g_2} = 0.05$, ${\beta _1} = {\beta _2} = 0.7$.}\label{fig2}
\end{figure}
\indent The ultra-sensitive vibrational resonance caused by small disturbances does not only mean that the high-frequency signal must be particularly weak, but also mean that a very small change in the high-frequency signal can cause the phenomenon. Actually, the amplitude of the high-frequency signal can vary over a large range, as shown in Fig.~3. Small disturbances in the amplitude and frequency of the high-frequency signal can cause significant fluctuations in the corresponding response amplitude over a large range. Further, from the figure, we also find that the ultra-sensitive vibrational resonance caused by the high-frequency signal is very complex. In addition to the presence of ultra-sensitive vibrational resonance in most areas, there are also very obvious resonant regions and non-resonant regions in the figure. Of course, the ultra-sensitive vibrational resonance does not exist in such areas. Based on subsequent analysis, we can further find that in these obvious resonant and non-resonant regions, the system exhibits a periodic response or a double periodic response, while in other ultra-sensitive resonant regions, the system response is chaotic and the response amplitude versus the controlled parameters is highly fractalized.\\

\begin{figure}[h!]
\centering
\includegraphics[width=5in]{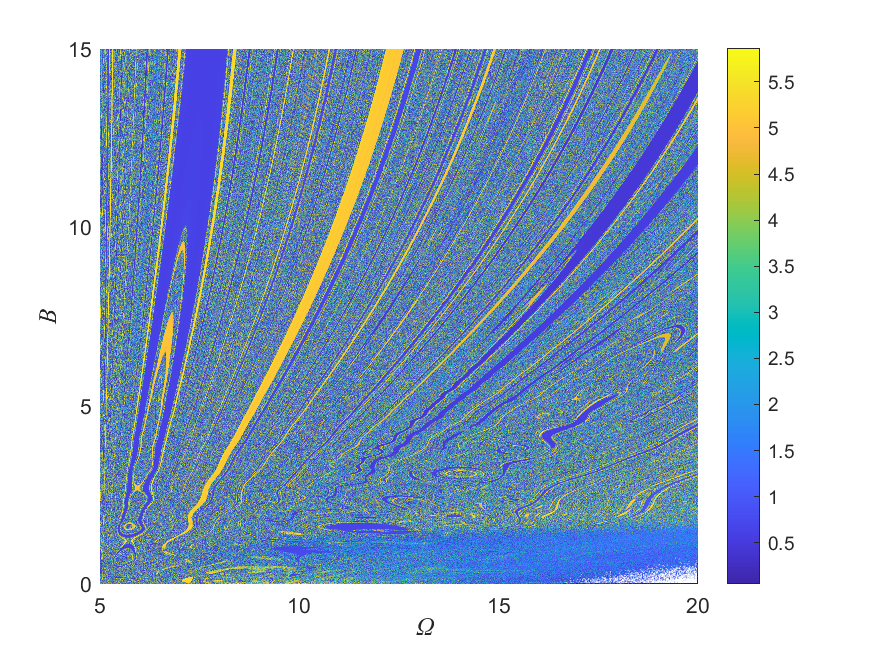}
\parbox{17cm}{\small {\bf FIG.~3.} The ultra-sensitive vibrational resonance of $x_1$ at the frequency $\omega$ caused by a small disturbance in large ranges of $B$ and $\Omega$. The color bar displays the value of $Q_1$. The parameters are $A = 0.25$, $\omega = 1$, ${\gamma _1} = {\gamma _2} = 0.05$, $\omega _1^2 = \omega _2^2 = 1$, ${g_1} = {g_2} = 0.05$, ${\beta _1} = {\beta _2} = 1$. Both the steps of $B$ and $\Omega$ are set as $1 \times 10^{-2}$.}\label{fig3}
\end{figure}
The ultra-sensitive vibrational resonance of $x_1$ not only occurs at the excitation frequency $\omega$ but also at many other nonlinear frequency components, as displayed in Fig.~4. It indicates that the ultra-sensitive vibrational resonance contains the nonlinear vibrational resonance. Moreover, in previous works, the nonlinear vibrational resonance occurs usually at the subharmonic frequency components, the subharmonic frequency components, or the combined frequency components. Due to the chaotic property of the response, the nonlinear resonance herein, i.e., the ultra-sensitive vibrational resonance  corresponding to a chaotic response occurs at a wide range of frequencies. The continuity of the spectrum $X_1(f)$ in Fig.~2 (b) can also illustrate this fact.\\
\begin{figure}
\centering
\includegraphics[width=6in]{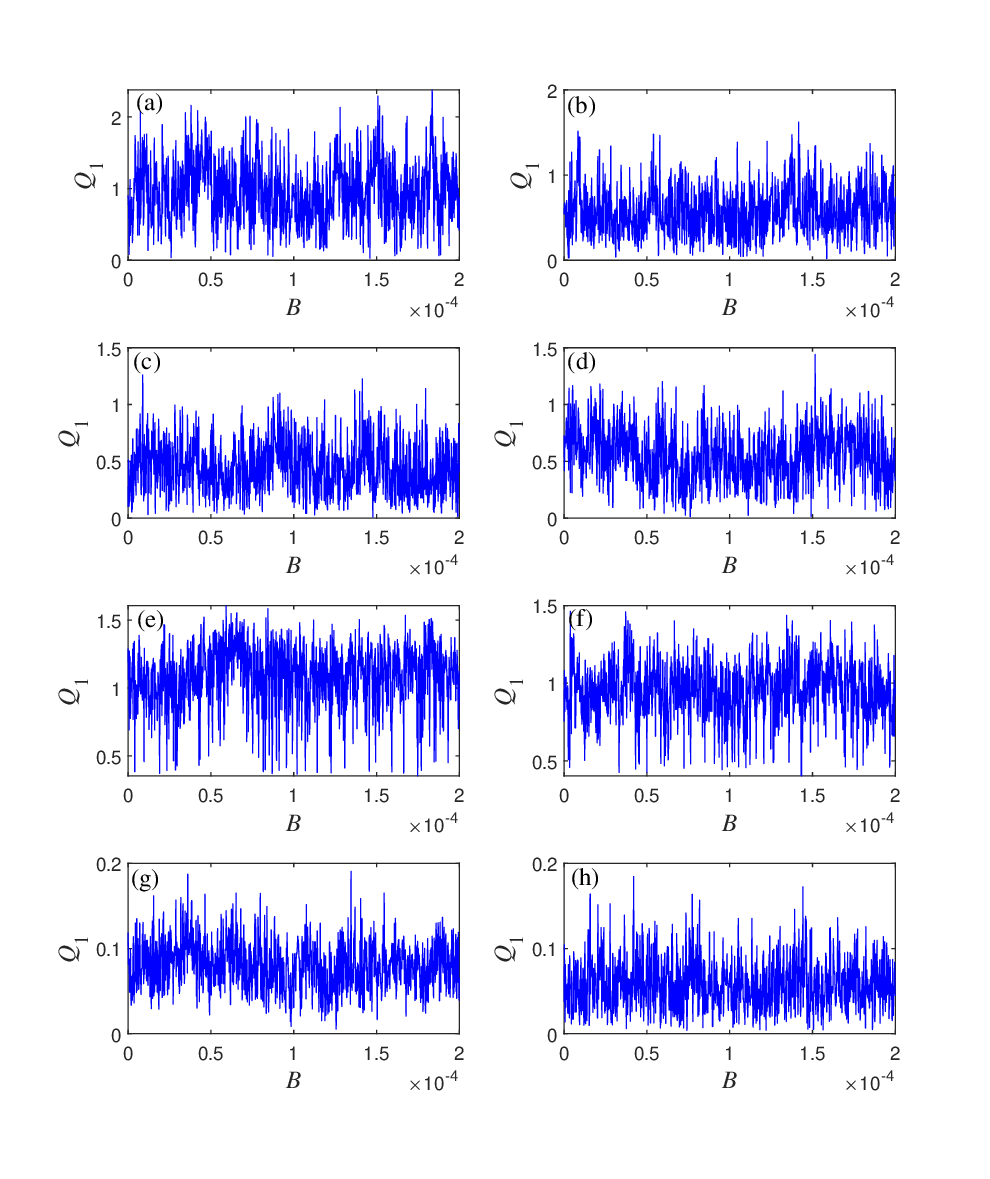}
\parbox{17cm}{\small {\bf FIG.~4.} A very small disturbance of the amplitude of the high-frequency signal induces ultra-sensitive vibrational resonance of $x_1$ at some nonlinear frequency components. In the left column, ${g_2} = 0.05$, while in the right column, ${g_2} = 0.3$. ${Q_1}$ is calculated at $m\omega$, in (a) and (b) $m=1/3$, in (c) and (d) $m=1/2$, in (e) and (f) $m=1$, in (g) and (h) $m=2$. The parameters are $A = 0.25$, $\omega  = 1$, $\Omega  = 10$, ${\gamma _1} = {\gamma _2} = 0.05$, $\omega _1^2 = \omega _2^2 = 0.8$, ${g_1} = 0.05$, ${\beta _1} = {\beta _2} = 0.7$, and the step of $B$ is $\Delta B=2 \times 10^{-7}$.}\label{fig4}
\end{figure}

\indent The ultra-sensitive vibrational resonance not only occurs on $x_1$, but also on $x_2$, even if there is no low-frequency signal acting on $x_2$, as shown in Fig.~5. This is due to the combined effect of the system coupling and the chaotic response. When the coupling strength is relatively large, such as $g_2=0.3$, the response amplitudes at some nonlinear frequencies are also relatively large, as illustrated in Figs.~5(b), (d), (f) and (h). The coupling is an important factor to cause strong nonlinear response of the variable $x_2$ without excitation.\\
\begin{figure}
\centering
\includegraphics[width=6in]{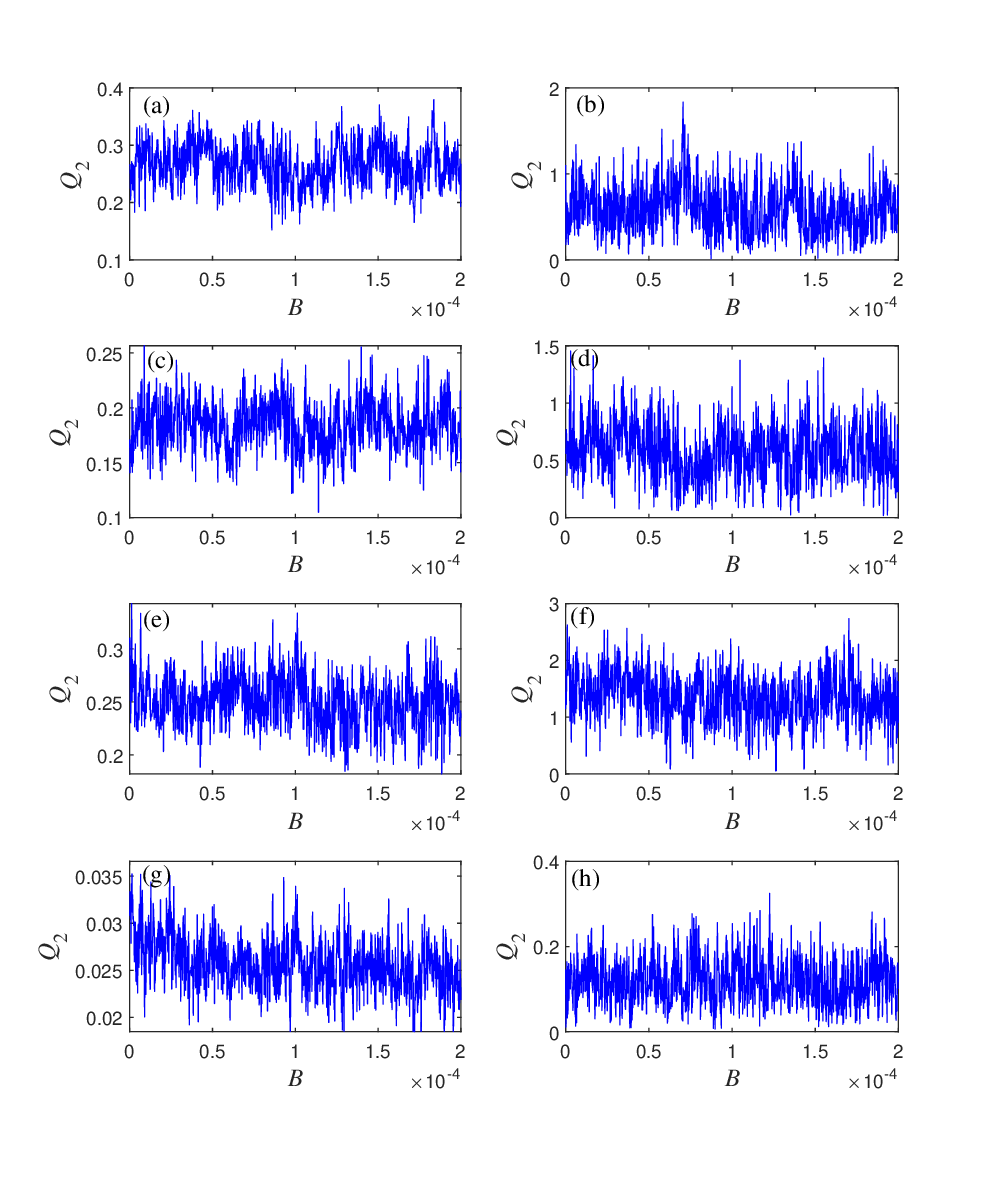}
\parbox{17cm}{\small {\bf FIG.~5.} A very small disturbance of the amplitude of the high-frequency signal induces ultra-sensitive vibrational resonance of $x_2$ at some nonlinear frequency components. In the left column, ${g_2} = 0.05$, while in the right column, ${g_2} = 0.3$. ${Q_2}$ is calculated at $m\omega$, in (a) and (b) $m=1/3$, in (c) and (d) $m=1/2$, in (e) and (f) $m=1$, in (g) and (h) $m=2$. The parameters are $A = 0.25$, $\omega  = 1$, $\Omega  = 10$, ${\gamma _1} = {\gamma _2} = 0.05$, $\omega _1^2 = \omega _2^2 = 0.8$, ${g_1} = 0.05$, ${\beta _1} = {\beta _2} = 0.7$, and the step of $B$ is $\Delta B=2 \times 10^{-7}$.}\label{fig5}
\end{figure}

\subsection{Ultra-sensitive vibrational resonance induced by a small initial condition disturbance}
A small disturbance can cause ultra-sensitive vibrational resonance not only when acting on the high-frequency signal, but also when acting on the initial conditions of the numerical simulation processes, as presented in Fig.~6. We choose the value of $B$ as a small value corresponding to the peak of the curve when the ultra-sensitive vibrational resonance occurs in Fig.~1. The initial conditions for numerical simulation of Eq.~(1) are labeled as $x_1(0)=x_{10}$, $x_2(0)=x_{20}$. In addition to the ultra-sensitive vibrational resonance, there are also multiple resonant and non-resonant regions in the figure. Both in Fig.~6 (a) and Fig.~6(b), the plot of $Q_1$ presents a highly fractalized characteristics. The difference in the system response caused by the initial conditions can be said to be vastly different. In such systems, the important impact of initial conditions on the system response cannot be ignored, which is different from previous extensive vibrational resonance research equations.\\
\begin{figure}
\centering
\includegraphics[width=7.2in]{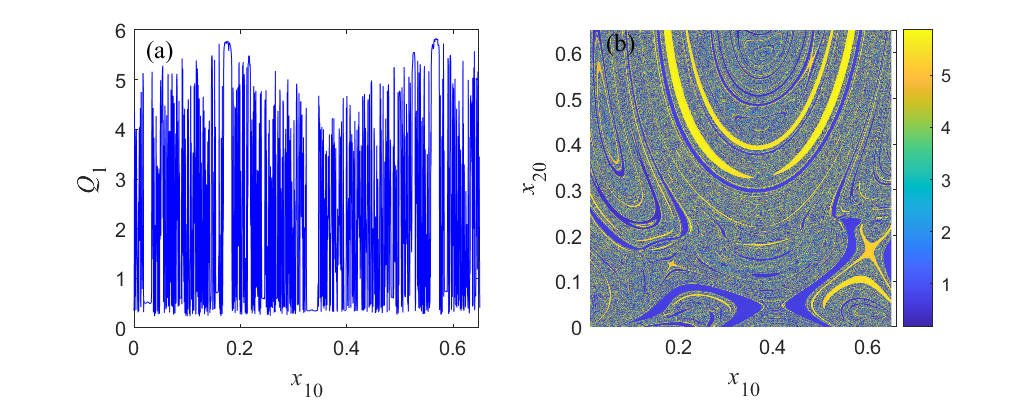}
\parbox{17cm}{\small {\bf FIG.~6.} The ultra-sensitive vibrational resonance of $x_1$ at the frequency $\omega$ caused by a small disturbance in the initial conditions $x_{10}$ and $x_{20}$. (a) $Q_1$ versus $x_{10}$, herein, $x_{10}=0:0.0005:0.65$, $x_{20}=0:0.0005:0.65$. (b) $Q_1$ versus $x_{10}$ and $x_{20}$. The color bar displays the value of $Q_1$. Both the calculation step of $x_{10}$ and $x_{20}$ is set as $5 \times 10^{-4}$. Other parameters are $A = 0.25$, $\omega  = 1$, $B = 0.0012$, $\Omega  = 10$, ${\gamma _1} = {\gamma _2} = 0.05$, $\omega _1^2 = \omega _2^2 = 1$, ${g_1} = {g_2} = 0.05$, ${\beta _1} = {\beta _2} = 1$.}\label{fig6}
\end{figure}
\section{Transformation between ultra-sensitive vibrational resonance and conventional vibrational resonance}
There is a phenomenon of ultra-sensitive vibrational resonance in nonlinear coupled systems, as mentioned in this paper, as well as conventional vibrational resonance, as described in [51]. What causes both conventional and ultra-sensitive vibrational resonances to occur in the system? This section will uncover the mystery to find out what factors are causing the system response to transform from the conventional vibrational resonance to ultra-sensitive vibrational resonance.\\
\indent From the curves in the left column of Fig.~7, we gradually increase the damping coefficient from a small value, and the ultra-sensitive vibrational resonance undergoes a process from appearing to disappearing. The ultra-sensitive vibrational resonance is transformed into the conventional vibrational resonance. Similarly, from the curves in the right column of Fig.~7, the vibrational resonance undergoes the same transformation process. It indicates that the damping coefficient will affect the occurrence of the ultra-sensitive vibrational resonance. When the damping coefficient is small, the ultra-sensitive vibrational resonance is more likely to occur. When the damping coefficient is large, this phenomenon will disappear. Furthermore, by comparing the curves in the left column with those in the right column, it is found that the small coupling force $g_1$ can lead to a stronger ultra-sensitive vibrational resonance. In addition, the smaller the value of $g_1$, the smaller the value $B$ that causes ultra-sensitive vibrational resonance, and the wider the region of ultra-sensitive vibrational resonance.\\
\begin{figure}
\centering
\includegraphics[width=6in]{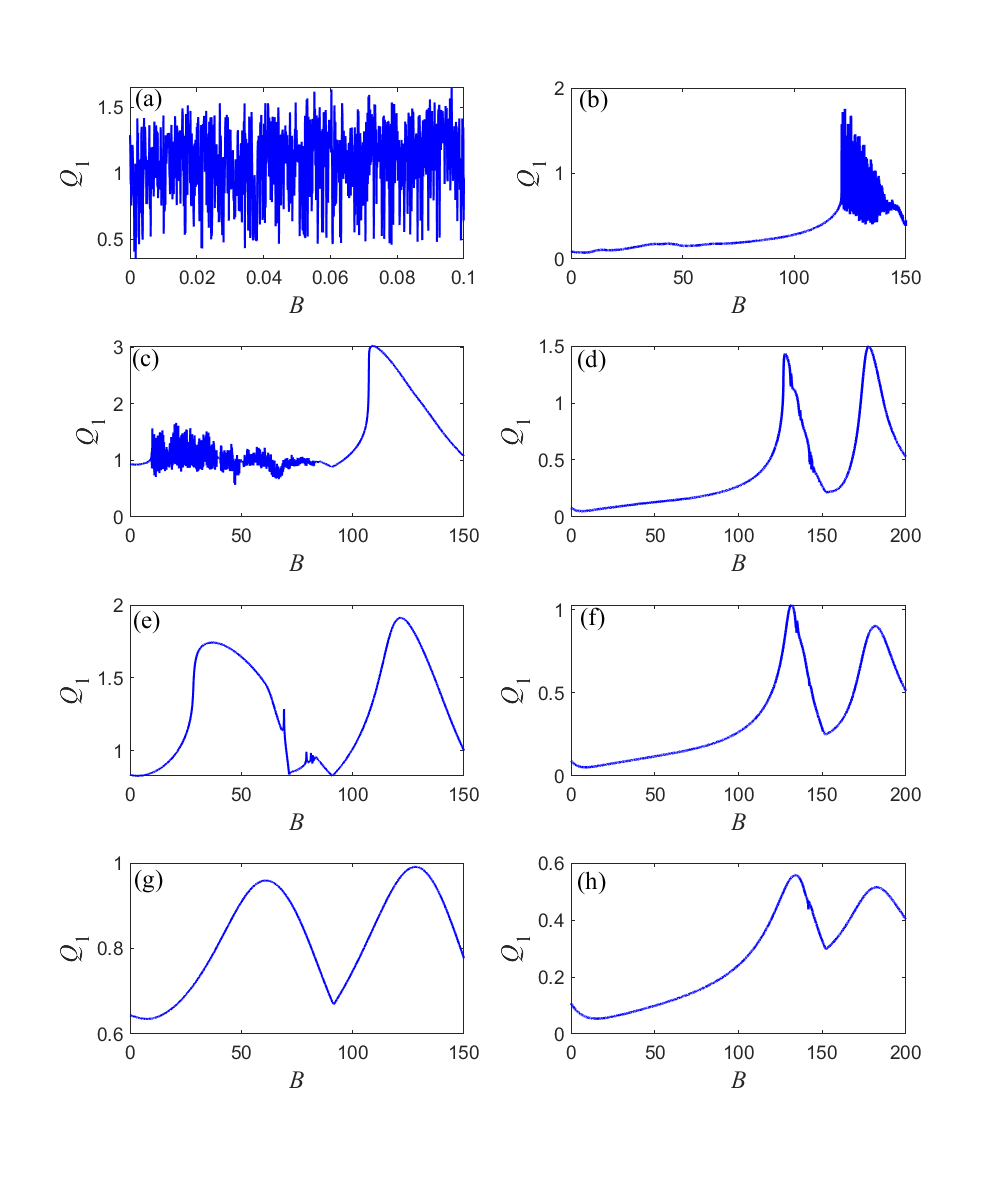}
\parbox{17cm}{\small {\bf FIG.~7.} The response amplitude of $x_1$ at the frequency $\omega$ versus the amplitude of the high-frequency signal presents a transformation from the ultra-sensitive vibrational resonance to the conventional vibrational resonance due to the change of the damping coefficient $\gamma_1$. In the left column, ${g_1} = {g_2} = 0.05$, while in the right column ${g_1} = {g_2} = 0.8$, in (a) and (b) $\gamma_1  = \gamma_2 =0.05 $, in (c) and (d) $\gamma_1  = \gamma_2 =0.13$, in (e) and (f) $\gamma_1  = \gamma_2 = 0.2$, in (g) and (h) $\gamma_1  = \gamma_2 = 0.4$. Other parameters are $A = 0.25$, $\omega  = 1$, $\Omega  = 10$, $\omega _1^2 = \omega _2^2 = 0.8, {\beta _1} = {\beta _2} = 0.7$.}\label{fig7}
\end{figure}
\indent To further investigate the influences of the system parameters on the vibrational resonance pattern presented in the system response, we provide Fig.~8, mainly focusing on the effect of the coupling strength. For the curves in the left column, we still select a relatively small damping coefficient $\gamma_1$. As the coupling strength increases, the zone of the ultra-sensitive vibrational resonance gradually moves to the right, and the fluctuation degree of the resonance gradually weakens, and the resonant region also gradually decreases. For the curves in the right column, we use a relatively large damping coefficient $\gamma_1$. When the coupling strength increases gradually, the response amplitude $Q_1$ versus the signal amplitude $B$ always presents the conventional vibrational resonance. Figure~8 once again illustrates that a small damping coefficient and a coupling strength are more likely to cause ultra-sensitive vibrational resonance. Otherwise, the system response may exhibit the conventional vibrational resonance pattern. As the damping system and the coupling strength increases, the system response pattern achieves a transformation from ultra-sensitive vibrational resonance to the conventional vibrational resonance.\\
\indent Although we only discuss the influences of $\gamma_1$ and $g_1$ in this section, we will have similar results for other system parameters.\\
\begin{figure}[h!]
\centering
\includegraphics[width=5.6in]{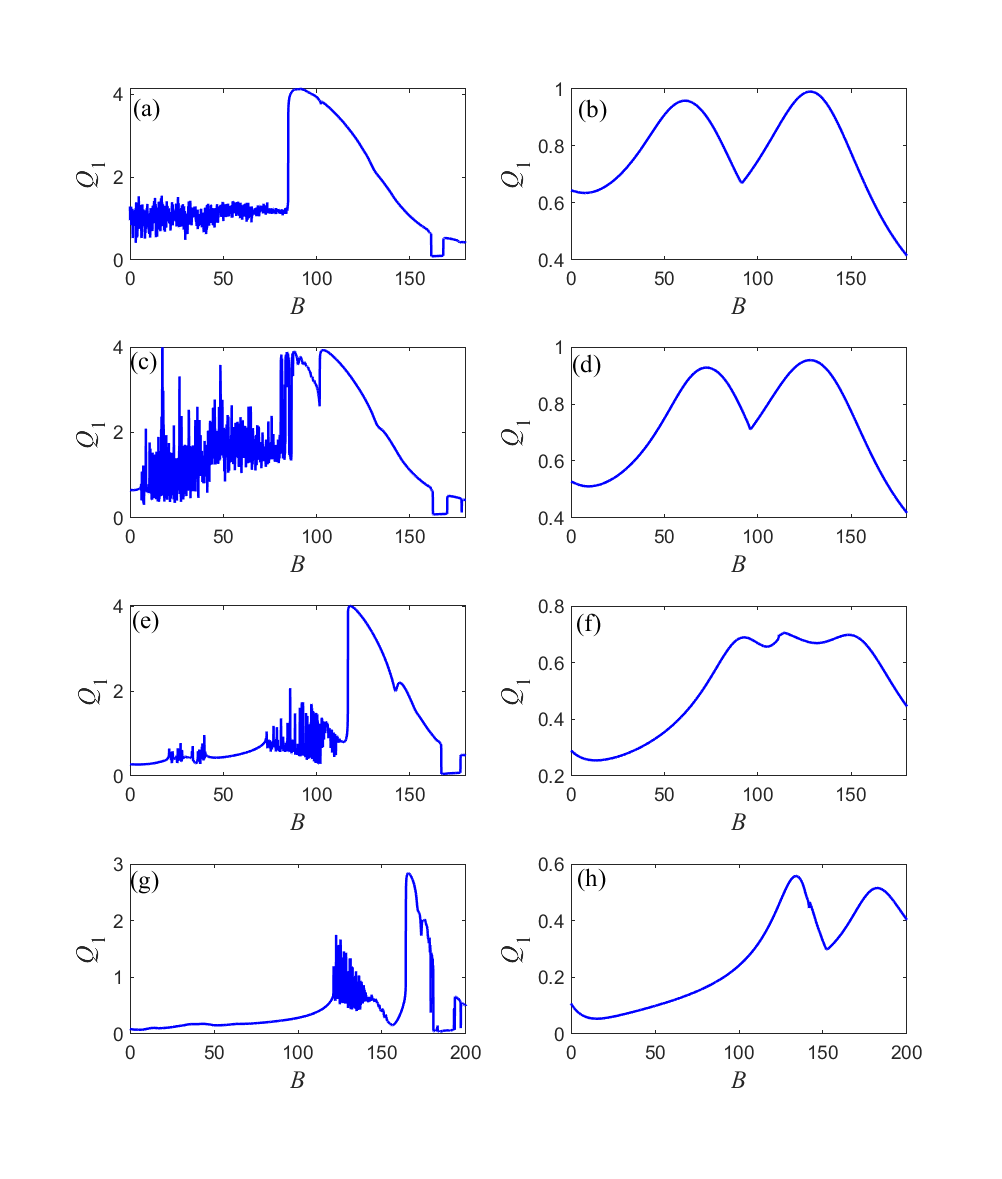}
\parbox{17cm}{\small {\bf FIG.~8.} The response amplitude of $x_1$ at the frequency $\omega$ versus the amplitude of the high-frequency signal presents a transformation from the ultra-sensitive vibrational resonance to the conventional vibrational resonance due to the change of the coupling strength $g_1$. In the left column, ${\gamma_1} = {\gamma_2} =0.05$, while in the right column ${\gamma_1} = {\gamma_2} = 0.4$, in (a) and (b) ${g_1} = {g_2} = 0.05$, in (c) and (d) ${g_1} = {g_2} = 0.1$, in (e) and (f) ${g_1} = {g_2} = 0.3$, in (g) and (h) ${g_1} = {g_2} =0.8$. Other parameters are $A = 0.25$, $\omega = 1$, $\Omega  = 10$, $\omega _1^2 = \omega _2^2 = 0.8$, ${\beta _1} = {\beta _2} = 0.7$.}\label{fig8}
\end{figure}
\section{Theoretical analysis for ultra-sensitive vibrational resonance}
To further understand the ultra-sensitive vibrational resonance phenomenon discussed above, we present the phase diagram and the Poincar\'{e} section in Fig.~9. At first, we let $B=0$. In other words, when there is no high-frequency excitation, as shown in Figs.~9(a) and (b), the response presents chaotic behavior. Due to the sensitivity of the chaotic response to a small disturbance, when a weak high-frequency disturbance is added, the system response may undergo a significant change, resulting in the ultra-sensitive vibrational resonance. In Figs.~9(c) and (d), when a very small high-frequency disturbance is added, $B=0.0012$, the response of the system still presents chaotic behavior. Corresponding to Fig.~1, when $B=0.0012$, the response amplitude $Q_1$ will achieve at one of the peaks.\\
\indent In fact, we can infer that the same principle can also explain the ultra-sensitive vibrational resonance caused by the initial conditions $x_1(0)$ and $x_2(0)$. This is determined by the sensitivity of the chaotic response to the initial disturbances. As a result, the presence of a chaotic response in the system is the reason for determining whether ultra-sensitive vibrational resonance occurs at the excitation frequency and some other frequencies.
\begin{figure}[h!]
\centering
\includegraphics[width=7in]{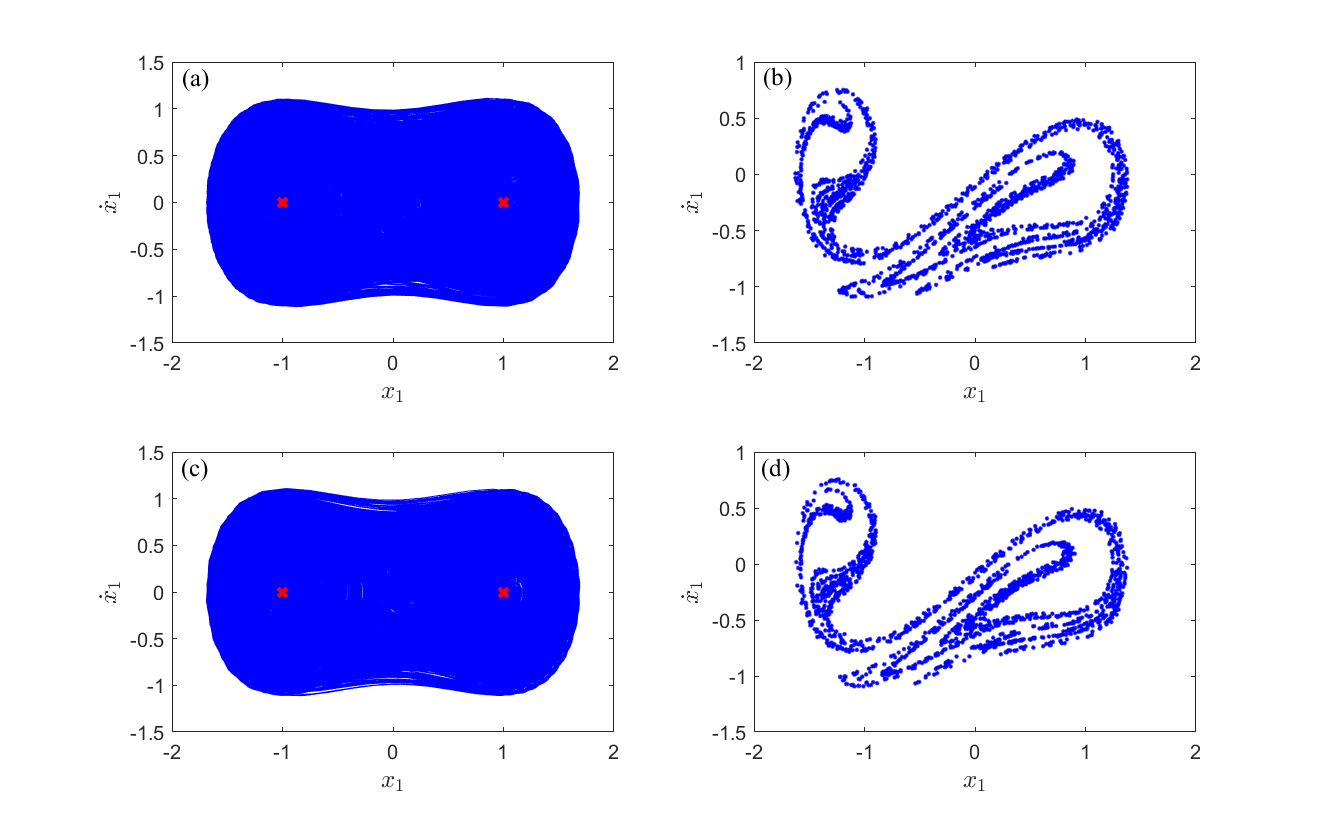}
\parbox{17cm}{\small {\bf FIG.~9.} The phase space diagram (left column) and the Poincar\'{e} section (right column) of the response. In (a) and (b), $B=0$; in (c) and (d), $B=0.0012$. Other parameters are $A = 0.25$, $\omega = 1$, $\Omega = 10$, ${\gamma _1} = {\gamma _2} = 0.05$, $\omega _1^2 = \omega _2^2 = 0.8$, ${g_1} = {g_2} = 0.05$, ${\beta _1} = {\beta _2} = 0.7$, $x_{10} = x_{20} = 0$.}\label{fig9}
\end{figure}

\indent As we have mentioned in Fig.~7, the damping coefficient is an important factor causing the ultra-sensitive vibrational resonance. In Fig.~10, we present the phase space diagrams and the corresponding Poincar\'{e} sections for different values of the damping coefficient. The phase space diagram presents chaotic motion in Figs.~10(a), (b), periodic doubling motion in Figs.~10(c), (d), periodic motion in Figs.~10(e) and (f), respectively. Once again, it shows that a change of the damping coefficient causes the system response to exhibit various states, which is the reason for the existence of different vibrational resonance patterns in the system.
\begin{figure}[h!]
\centering
\includegraphics[width=6in]{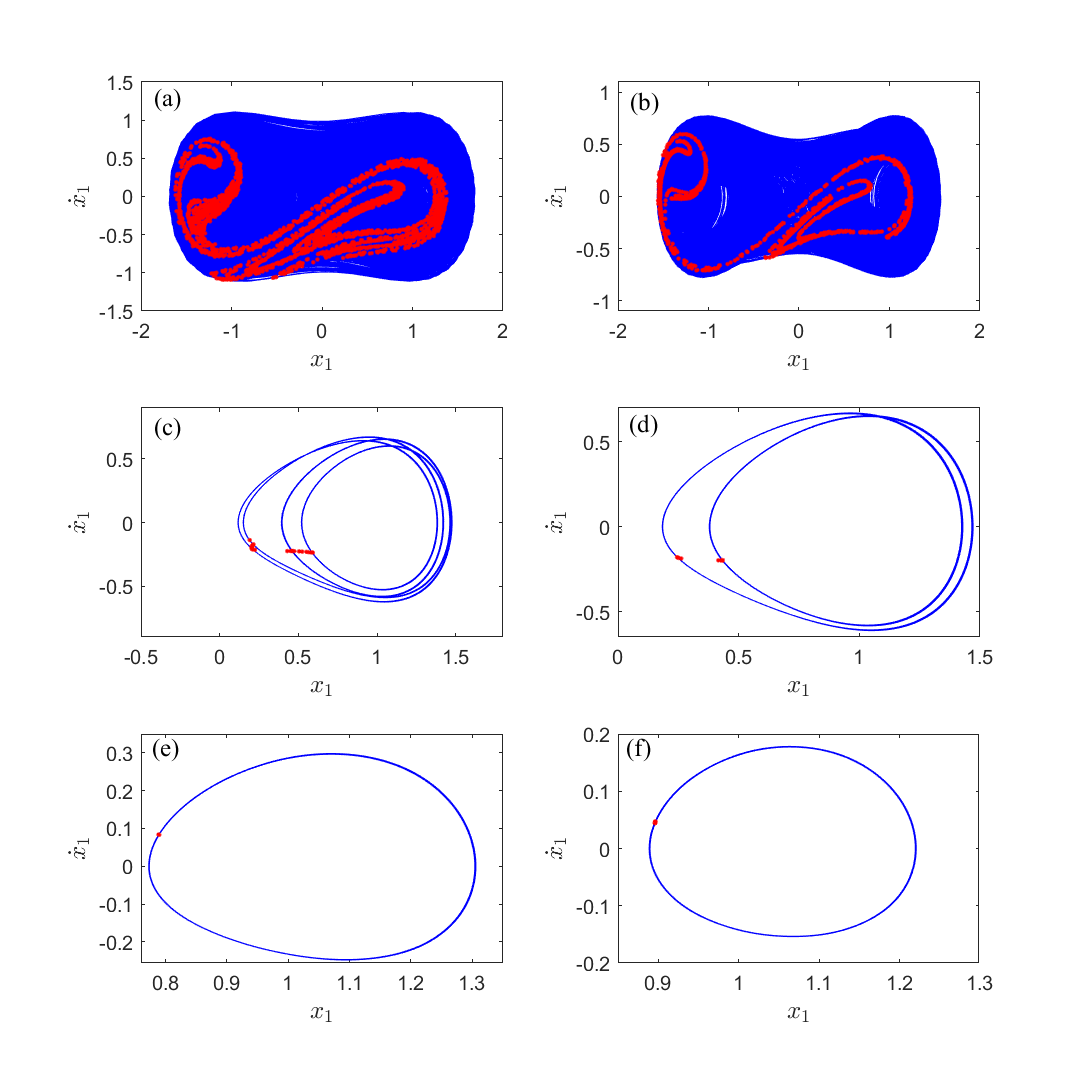}
\parbox{17cm}{\small {\bf FIG.~10.} The phase space diagrams for different values of the damping coefficient $\gamma_1$. The discrete points are the Poincar\'{e} cross-section. The parameters are $A = 0.25$, $\omega  = 1$, $B = 0.0012$, $\Omega = 10$, $\omega _1^2 = \omega _2^2 = 0.8$, ${g_1} = {g_2} = 0.05$, ${\beta _1} = {\beta _2} = 0.7$, and ${\gamma _1} = {\gamma _2}$ = $0.05$, $0.11$, $0.16$, $0.17$, $0.4$, $0.7$ from (a) to (f).}\label{fig10}
\end{figure}
\section{Conclusions}
In this paper, we investigate the ultra-sensitive vibrational resonance at both the excitation frequency and other nonlinear frequencies in a coupled system excited by high-frequency and low-frequency signals. We have found two different kinds of ultra-sensitive vibrational resonance induced by small disturbances. The first one is induced by a high-frequency disturbance. The second one is induced by a disturbance in the initial conditions.\\
\indent The main conclusions are summarized and listed as follows.\\
(1) The ultra-sensitive vibrational resonance is a transient phenomenon, which means that an irregular state stabilizes in a periodic state after a long transient.\\
(2) The ultra-sensitive vibrational resonance occurs not only at the excitation frequency, but also at a wide range of nonlinear frequencies.\\
(3) Due to the coupling between the two variables, the response amplitude $Q_2$ exhibits ultra-sensitive vibrational resonance phenomenon even when there is no characteristic low-frequency signal acting on $x_2$.\\
(4) In addition to being affected by a high-frequency disturbance, the ultra-sensitive vibrational resonance also occurs when the initial conditions are slightly disturbed.\\
(5)  The damping coefficient and the coupling strength are the key factors leading to the ultra-sensitive vibrational resonance. They also influence the transformation of vibrational resonance patterns from the ultra-sensitive one to the conventional one. Especially for small damping coefficient and small coupling strength, it is easy to induce ultra-sensitive vibrational resonance. With the increase of the damping coefficient and coupling strength, the vibrational resonance pattern can be transformed from the ultra-sensitive vibrational resonance to the conventional vibrational resonance.\\
\indent We believe that the ultra-sensitive vibrational resonance is worth studying, as a way to better understand  the complex dynamical response of nonlinear systems, as well as explore its applications in science and engineering.
\section*{Acknowledgements}
We appreciate useful discussions with Prof. Chenggui Yao in Jiaxing University and Dr. Jinjie Zhu in Nanjing University of Aeronautics and Astronautics.\\
\indent The project was supported by the National Natural Science Foundation of China (Grant No. 12072362), the Postgraduate Research \& Practice Innovation Program of Jiangsu Province (Grant No. KYCX23-2682),the Graduate Innovation Program of China University of Mining and Technology (Grant No. 2023WLJCRCZL103), the Priority Academic Program Development of Jiangsu Higher Education Institutions, the Spanish State Research Agency (AEI) and the European Regional Development Fund (ERDF, EU) under Project No. PID2019-105554GB-I00 (MCIN/AEI/10.13039/501100011033).

\end{document}